\documentclass[english,aps,reprint,showpacs,amsmath,amssymb]{revtex4-1}
\usepackage[T1]{fontenc}
\usepackage[latin9]{inputenc}
\usepackage{amsmath}
\usepackage{amssymb}
\usepackage{graphicx}
\usepackage{natbib}
\usepackage{babel}	
\usepackage[running,displaymath, mathlines]{lineno}
\begin{document}

\title{Role of Dimensionality and Size in Controlling the Drag Seebeck Coefficient
of Doped Silicon Nanostructures: A Fundamental Understanding}

\author{Raja Sen}
\email{rajasenphysics@gmail.com}

\author{Nathalie Vast}

\author{Jelena Sjakste}
\email{jelena.sjakste@polytechnique.edu}

\affiliation{Laboratoire des Solides Irradies, CEA/DRF/IRAMIS, Ecole Polytechnique,
CNRS, Institut Polytechnique de Paris, 91120 Palaiseau, France}

\begin{abstract}
In this theoretical study, we examine the influence of dimensionality,
size reduction, and heat-transport direction on the phonon-drag contribution to the Seebeck coefficient of silicon nanostructures. 
Phonon-drag contribution, which arises from the momentum transfer between out-of-equilibrium phonon populations and charge carriers, 
significantly enhances the thermoelectric coefficient. Our implementation of the
  phonon drag term accounts for the anisotropy of  nanostructures such as thin films and nanowires through the boundary- and momentum-resolved phonon lifetime.
 Our approach also takes into account the spin-orbit coupling which turns out to be crucial for hole transport.
 We reliably quantify the phonon drag contribution at various doping levels, temperatures, and nanostructure geometries
 for both electrons and holes in silicon nanostructures. Our results support the recent experimental findings,
showing that a part of phonon drag contribution survives in 100~nm silicon nanostructures.
\end{abstract}


\maketitle



The nanostructuring of semiconductors provides a viable route to enhance
the thermoelectric efficiency as compared to that of the bulk by
tuning the transport properties~\citep{He:2017,Dangic:2020,Yang:2021}.
Together with the electrical and  thermal conductivities,
the Seebeck coefficient -~which links the electrical current to the temperature gradient~- is a key physical quantity
characterizing the performance of thermoelectric materials.
For a nondegenerate semiconductor,
there are two contributions to the total Seebeck coefficient ($\mathrm{\mathit{S}^{tot}}$):
the diffusive ($\mathrm{\mathit{S}^{diff}}$) and the
phonon drag ($\mathrm{\mathit{S}^{drag}}$) contributions.
While the former comes from the diffusion of charge carriers under
a temperature gradient, the latter arises from the momentum transfer
between the out-of-equilibrium phonon populations and the
charge carriers~\citep{Gurevich:1945,Herring:1954,Cantrell_I:1987,Cantrell_II:1987,Gurevich:1989}.
Despite the impossibility to separately measure the $\mathrm{\mathit{S}^{diff}}$ and $\mathrm{\mathit{S}^{drag}}$
contributions, the important role played by the phonon drag has been recognized experimentally by the strong increase of $\mathrm{\mathit{S}^{tot}}$ at low temperatures in semiconductors, where anharmonicity is reduced
and out-of-equilibrium phonon populations are very
large~\citep{Frederikse:1953,Geballe:1954,Geballe:1955,Bentien:2007,Takahashi:2016,Jaoui:2020}. 

The individual contributions to $\mathrm{\mathit{S}^{tot}}$ can, however, be quantified theoretically
by means of models with effective parameters~\citep{Herring:1954,Mahan:2014,Battiato:2015}
or by \textit{ab initio} calculations~\citep{Zhou:2015,Fiorentini:2016,Protik:2020,Xu:2021,Protik:2022}.
For example, it has been recently shown by density functional theory (DFT)
that at $300$~K and at low electron doping ($10^{14}~\mathrm{cm^{-3}}$),
more than $30\%$ of $\mathrm{\mathit{S}^{tot}}$ in silicon comes from $\mathrm{\mathit{S}^{drag}}$.
Moreover, the relative contribution of $\mathrm{\mathit{S}^{drag}}$ with respect to $\mathrm{\mathit{S}^{tot}}$ 
increases even further at higher doping (and fixed temperature) or at (fixed doping and) temperatures lower
than $300$~K~\citep{Zhou:2015,Fiorentini:2016,Protik:2020,Xu:2021,Protik:2022}.
Nevertheless, downsizing a semiconductor to the sub-micron scale is expected to drastically reduce
the mean free path (MFP) of phonons and in consequence, the drag contribution to $\mathrm{\mathit{S}^{tot}}$.
Therefore, the Seebeck coefficient is expected to decrease monotonically with decreasing the size of the nanostructure~\citep{Frederikse:1953,Geballe:1954,Geballe:1955,Takahashi:2016}.

In spite of the great effort invested in studying the effect of nanostructuring
on $\mathrm{\mathit{S}^{tot}}$~\citep{Pokharel:2013,Wang:2013,Kockert:2019,Nadtochiy:2019,Cabero:2021},
no consensus has been reached about the role of $\mathrm{\mathit{S}^{drag}}$ at the nanoscale,
even in the case of silicon~\citep{Zhou:2015,Sadhu:2015,Salleh:2014,Boukai:2008,Fauziah:2020,Ryu:2010}.
Indeed, from the theoretical side,
 Zhou \textit{et al.}~\citep{Zhou:2015} have
pointed out that phonons contributing to $\mathrm{\mathit{S}^{drag}}$
have longer MFPs than those contributing to the lattice thermal conductivity
of Si. A consequence is that $\mathrm{\mathit{S}^{drag}}$ is
expected to be strongly suppressed at $1$~$\mu$m by the effect of size reduction~\citep{Zhou:2015}.
 In addition, in the experimental study of $\mathrm{\mathit{S}^{tot}}$,
 Sadhu \textit{et al.}~\citep{Sadhu:2015} have concluded
that the $\mathrm{\mathit{S}^{drag}}$ component vanishes completely
in Si nanowires having a characteristic length smaller than $100$
nm. However, these studies contradict the findings of other experimental works which have suggested
that the drag contribution in
Si ultrathin films~\citep{Salleh:2014}, nanowires~\citep{Boukai:2008,Fauziah:2020},
and nanoribbons~\citep{Ryu:2010} does not vanish.
The situation is further complicated by the possible presence of various competing effects at the nanoscale,
such as the energy filtering effect induced by defects, which can lead to an
increase of the total Seebeck coefficient~\cite{Zide:2006,Bennett:2015,Liang:2017}.
Thus, to disentangle the intricate effects that govern the magnitude of $\mathrm{\mathit{S}^{tot}}$ at the nanoscale,
a detailed understanding of the dependence of $\mathrm{\mathit{S}^{drag}}$ on the
dimensionality and size of nanostructures is necessary and  can only be achieved through theory.

\begin{figure}[t!]
\includegraphics[width=8.7cm]{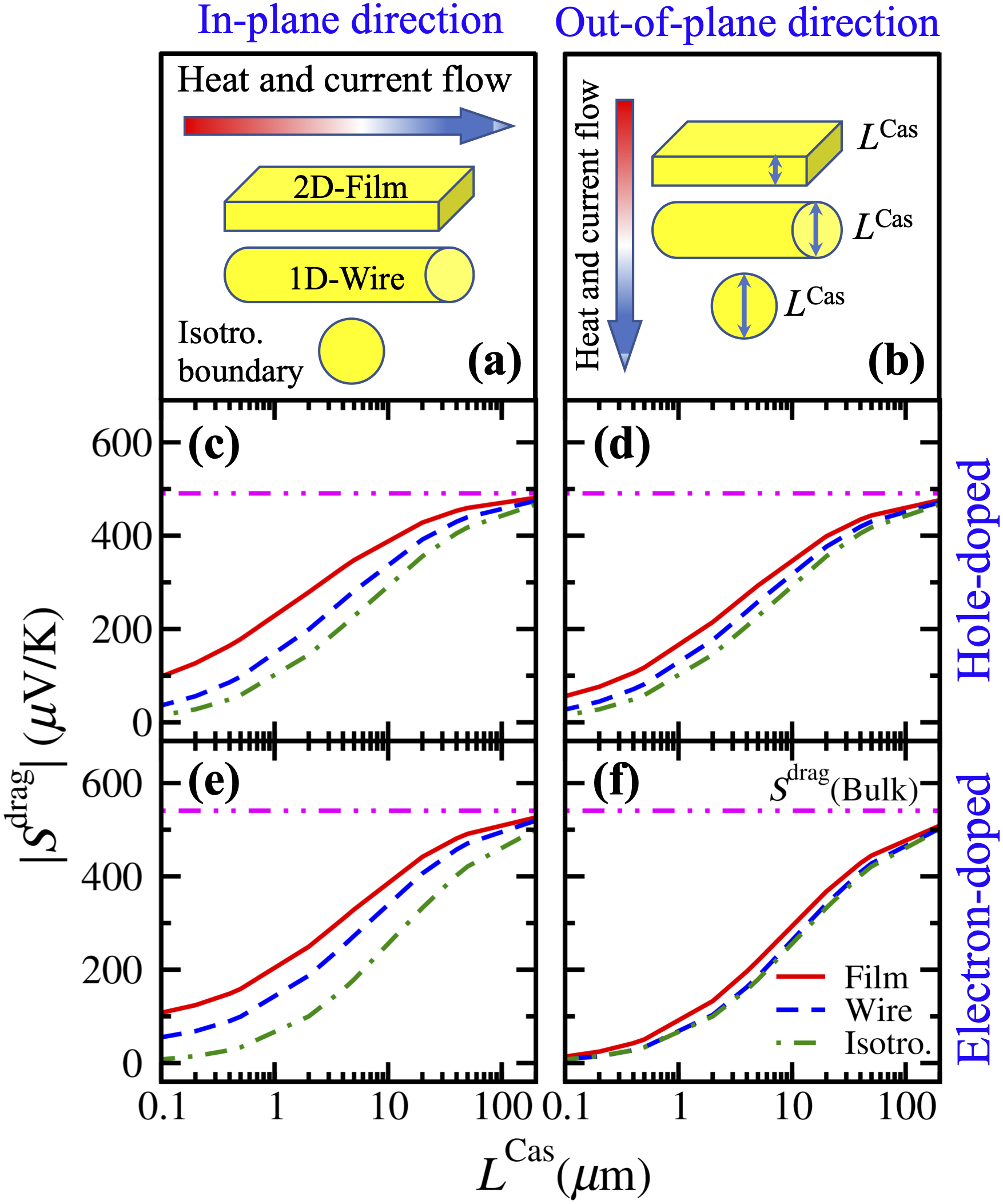}
\caption{Panel a (b): Schema of the in-plane (out-of-plane) direction of heat and current flows as a function
of the nanostructure dimensionality: 2D-Film, 1D-wire, and the isotropic boundary. Panel c (e): Variation of hole (electron) phonon drag Seebeck coefficient as a function of the Casimir length ($L^{\mathrm{Cas}}$) in the in-plane direction at $300$~K and $10^{14}~\mathrm{cm^{-3}}$ doping concentrations. Panel d (f): Same as panel c (e) in the out-of-plane
direction. \label{fig:1}}
\end{figure}

In this paper, we report the results of the systematic investigation of the influence of both size reduction and dimensionality on the phonon drag Seebeck coefficient
of electron- and hole-doped silicon, by solving the coupled linearized Boltzmann transport equation (BTE)
for charged carriers and for phonons, in combination with a fully \textit{ab initio} description
of the carrier-phonon interaction~\citep{Murphy-Armando:2006,Calandra:2010,Bernardi:2016,Giustino:2017b,Sjakste:2018a,Ponce:2020,Brunin:2020}.
The coupling of the BTEs enables us to include in particular the effect of the out-of-equilibrium phonon populations
which arise in presence of the temperature gradient.
At variance with previous ones, the computational approach implemented in this work allows to account
for the anisotropy of  phonon scattering by nanostructure boundaries in the calculations
of phonon drag Seebeck coefficient.
In our work, the role of anisotropy and dimensionality of boundary scattering
 has  been studied
by considering two-dimensional (2D) nanofilms and one-dimensional (1D) nanowires,
as well as isotropic boundaries
 (Fig.~\ref{fig:1}, panel~a).
The direction-resolved out-of-equilibrium phonon populations have been determined as a function of the
nanostructure size, with the aim of quantifying the effect of the transport-direction-dependent phonon-boundary scattering
in the phonon drag contribution. Moreover we have included the effect of spin-orbit coupling
on the $\mathrm{\mathit{S}^{tot}}$ for holes, an effect which has been neglected so far for the Seebeck coefficient.

The electrical and heat currents produced by a temperature gradient
experience a mutual drag via the interaction between charge
carriers and phonons. This means that, in principle, the carrier-phonon scattering terms
that govern the BTE for charge carriers and sometimes play a role in the BTE for phonons, depend both on the charge-carrier out-of-equilibrium distribution functions, $f_{\mathrm{n}\mathbf{k}}$, and on the out-of-equilibrium phonon populations $n_{\mathbf{q}\nu}$, where $\mathrm{n},~\mathbf{k},~\nu,~\mathbf{q}$ are respectively the electronic band index, wave vector, phonon mode index, and wave vector. However, the electron-phonon scattering terms in the phonon BTE has proven necessary only close to the degenerate semiconductor limit, e.g., for carrier concentrations
larger than $10^{19}~\mathrm{cm^{-3}}$ at 300~K in silicon~\cite{Zhou:2015}. In that case, a partial decoupling scheme~\cite{Cantrell_I:1987,Cantrell_II:1987} can be used, in which the electron-phonon scattering terms in the phonon BTE are made dependent on the (equilibrium) Fermi-Dirac distribution function
$f_{\mathrm{n}\mathbf{k}}^{0}$~\cite{Zhou:2015,Protik:2020,Xu:2021,Protik:2022}.
For low to moderate doping concentrations which do not exceed  $10^{19}~\mathrm{cm^{-3}}$, the effect of electron-phonon scattering on the phonon populations can be safely neglected~\cite{Mahan:2014,Fiorentini:2016,Macheda:2018}.
In the present study, we follow the latter approximation and obtain the Seebeck coefficient of silicon nanostructures, including the phonon drag mechanism, by solving the charge-carrier BTE in the relaxation time approximation:

\begin{linenomath}
\begin{align}
-\frac{\partial f_{\mathrm{n}\mathbf{k}}^{0}}{\partial\varepsilon_{\mathrm{n}\mathbf{k}}}\mathrm{\mathbf{v}_{n\mathbf{k}}}\cdot\left[\mathbf{E}e+\frac{\nabla_{\mathbf{r}}T}{T}\left(\varepsilon_{\mathrm{n}\mathbf{k}}-\mu\right)\right]\nonumber \\
-D_{\mathrm{n}\mathbf{k}}^{\mathrm{drag}}\left(g,\delta n\right) =& -\left(\frac{\partial f_{\mathrm{n}\mathbf{k}}}{\partial t}\right)_{coll}\label{eq:1}
\end{align}
\end{linenomath}
where $\mathbf{E}$ and $\nabla_{\mathbf{r}}T$ denote a small electric
field and the temperature gradient, $\mu$ is the chemical potential for holes (electrons),
$\mathbf{v}_{\mathrm{n}\mathbf{k}}$ and $\varepsilon_{\mathrm{n}\mathbf{k}}$
are respectively the charge-carrier group velocity and energy.
The term $\left(\frac{\partial f_{\mathrm{n}\mathbf{k}}}{\partial t}\right)_{coll}$
includes all of the collisions associated with carrier-phonon and
carrier-impurity scatterings.

The term $D_{\mathrm{n}\mathbf{k}}^{\mathrm{drag}}\left(g,\delta n\right)$
is the key quantity that describes the phonon drag mechanism, and can be written as~\citep{Fiorentini:2016,Macheda:2018}:

\begin{linenomath}
\begin{align}
D_{\mathrm{n}\mathbf{k}}^{\mathrm{drag}}\left(g,\delta n\right)= & \frac{2\pi}{\hbar}\sum_{\mathrm{m}\nu}\int\frac{d\mathbf{q}}{\Omega_{\mathrm{BZ}}}\left|g_{\mathrm{mn}\nu}\left(\mathbf{k},\mathbf{q}\right)\right|^{2}\nonumber \\
 & \times\left\{ \delta n_{\mathbf{q\nu}}\delta\left(\varepsilon_{\mathrm{n}\mathbf{k}}-\varepsilon_{\mathrm{m}\mathbf{k}+\mathbf{q}}+\hbar\omega_{\mathbf{q}\nu}\right)\right.\nonumber \\
 & +\left.\delta n_{\mathbf{-q\nu}}\delta\left(\varepsilon_{\mathrm{n}\mathbf{k}}-\varepsilon_{\mathrm{m}\mathbf{k}-\mathbf{q}}-\hbar\omega_{-\mathbf{q}\nu}\right)\right\} \nonumber \\
 & \times\left(f_{\mathrm{m}\mathbf{k}+\mathbf{q}}^{0}-f_{\mathrm{n}\mathbf{k}}^{0}\right)\label{eq:2}
\end{align}
\end{linenomath}
where $g_{\mathrm{mn}\nu}\left(\mathbf{k},\mathbf{q}\right)$ is the carrier-phonon
interaction matrix element and $\Omega_{\mathrm{BZ}}$ is the volume of the first Brillouin zone (BZ).
The (linearized) out-of-equilibrium phonon populations, $\delta n_{\mathbf{q}\nu}= n_{\mathbf{q}\nu} -  n_{\mathbf{q}\nu}^0$, are
expressed as:
\begin{equation}
\delta n_{\mathbf{q\nu}}=-\tau_{\mathbf{q\nu}}\frac{\nabla_{\mathbf{r}}T\cdot\mathrm{\mathbf{c}_{\mathbf{q\nu}}}}{k_{B}T^{2}}\hbar\omega_{\mathbf{q\nu}}n_{\mathbf{q\nu}}^{0}\left(1+n_{\mathbf{q\nu}}^{0}\right)\label{eq:3}
\end{equation}
and have been obtained by solving the phonon BTE in the single mode approximation with the D3Q code~\citep{Paulatto:2013,Fugallo:2013}.
Here, $\mathbf{c}_{\mathbf{q}\nu}$,
$\tau_{\mathbf{q}\nu}$, $\omega_{\mathbf{q}\nu}$, and $n_{\mathbf{q}\nu}^{0}$
are respectively the group velocity, lifetime, frequency, and the (equilibrium) Bose-Einstein phonon populations. 
We have used our in-house modified version of the Electron-Phonon Wannier (EPW) code~\citep{Ponce:2016} to solve Eq. \ref{eq:1}
(see Supplemental Material (SM) for implementation description
and computational details~\citep{SenNote:2023}).

Eqs.~\ref{eq:1} and ~\ref{eq:2} give rise to an electrical current in the same direction as the heat flow (Fig. \ref{fig:1}, panels~a~and~b). Indeed, although expressed in a form similar to standard electron-phonon scattering processes, the term
$D_{\mathrm{n}\mathbf{k}}^{\mathrm{drag}}\left(g,\delta n\right)$ is not a resistive process and contributes to the enhancement of the electrical current in the direction of the heat flow.
We obtain the diffusion contribution
when $\delta n_{\mathbf{q\nu}} = 0$, and thereby calculate $\mathrm{\mathit{S}^{drag}}$ as
$\mathrm{\mathit{S}^{tot}}|_{\delta n_{\mathbf{q\nu}} \neq 0} - \mathrm{\mathit{S}^{diff}}$.

In addition to the phonon-phonon and isotope scattering rates~\citep{Paulatto:2013,Fugallo:2013},
 the inverse of the phonon lifetime $(\tau_{\mathbf{q\nu}})^{-1}$
is also determined by the phonon-boundary scattering rate which, for nanostructures, is the crucial quantity that controls the magnitude of $D_{\mathrm{n}\mathbf{k}}^{\mathrm{drag}}$.
We have used the phonon momentum-resolved Casimir model
to determine the transport-direction-dependent phonon-boundary scattering
in nanostructures. The Casimir scattering rate for a phonon is given by~\citep{Markov:2016,Fauziah:2020}:
\begin{equation}
(\tau_{\mathbf{q\nu}}^{\mathrm{bound}})^{-1}=\left(\frac{1-p}{1+p}\right)\frac{|\mathbf{c}_{\mathbf{q}\nu}^{\mathrm{proj}}|}{L^{\mathrm{Cas}}}\label{eq:4}
\end{equation}
where the Casimir scattering length, $L^{\mathrm{Cas}}$, represents
the nanostructure size (Fig. \ref{fig:1}, panels~a~and~b). The specularity, $p$, ranges from 0 to 1 for completely diffusive
to completely specular scattering, respectively. The velocity, $|\mathbf{c}_{\mathbf{q}\nu}^{\mathrm{proj}}|$, is
the phonon group velocity $\mathbf{\mathbf{c}_{\mathbf{q}\nu}}$
projected on the direction(s) in which the phonon transport is limited by the boundaries (see SM).
It should be mentioned that while the anisotropy of the
boundary scattering has been taken into account in several theoretical studies of
lattice thermal conductivity~\citep{McGaughey:2011,Ma:2012,Jeong:2012,Markov:2016},
only the isotropic boundary (Fig. \ref{fig:1}, panel~a) has been
considered so far in the studies of the phonon drag effect~\citep{Fauziah:2020,Li:2022}.
The  case of anisotropic phonon-boundary scattering presents
a computational challenge, due to the reduced symmetry
of the term $D_{\mathrm{n}\mathbf{k}}^{\mathrm{drag}}\left(g,\delta n\right)$ of Eq.~\ref{eq:2}
in presence of the anisotropic phonon lifetime $\tau_{\textbf{q}\nu}$.
This challenge can be overcome by performing the calculation of $D_{\mathrm{n}\mathbf{k}}^{\mathrm{drag}}\left(g,\delta n\right)$
without making use of crystal symmetry considerations and employing an additional $\mathbf{q}$-points filtering scheme (see SM).

\begin{figure}[t!]
\includegraphics[width=8.7cm]{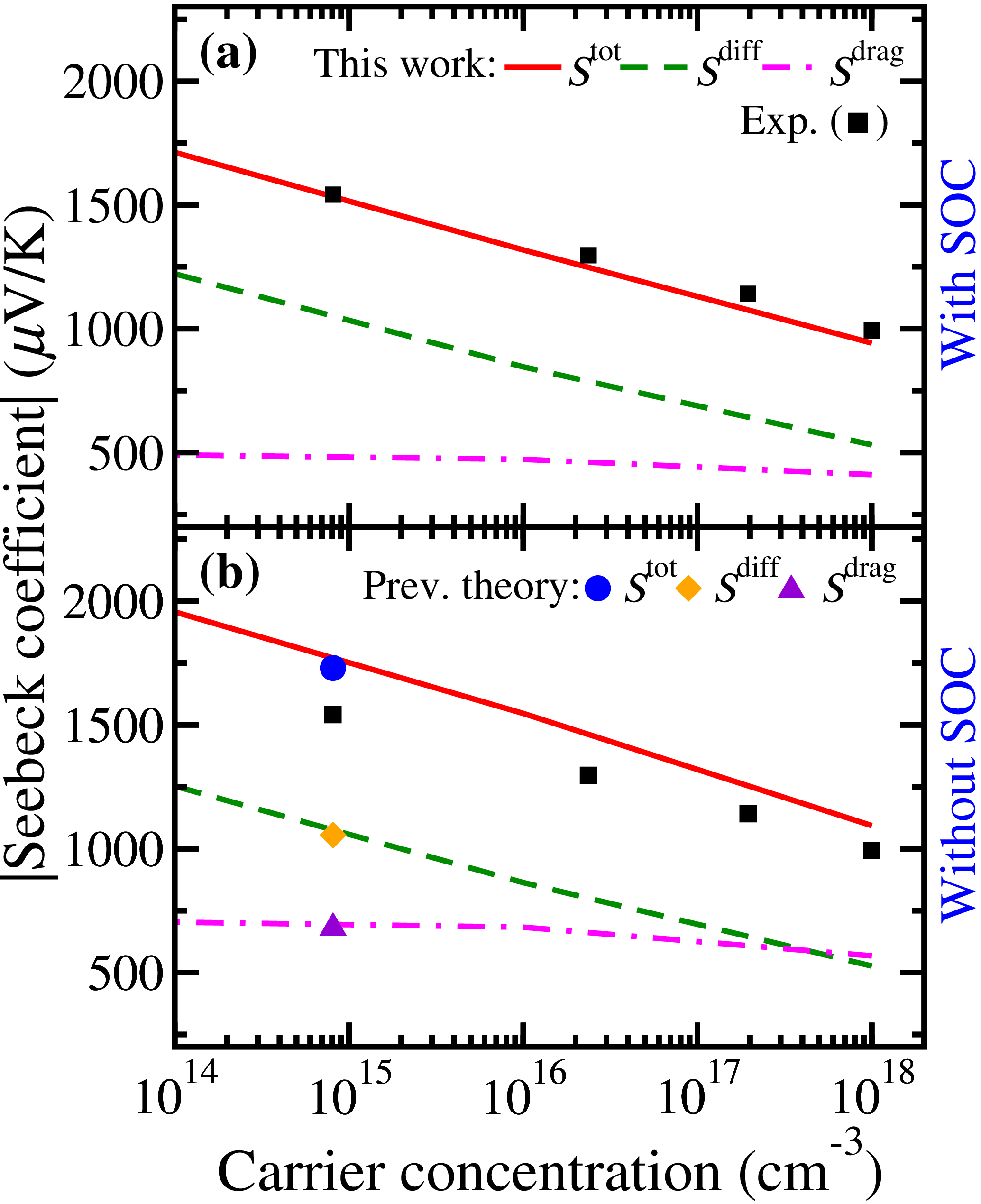}
\caption{Calculated Seebeck coefficient of hole-doped bulk
silicon (a) with and (b) without SOC interaction as a function of
the carrier concentration at 300~K. Solid line: $S^{\mathrm{tot}}$,
Dashed line: $S^{\mathrm{diff}}$, and Dot-dashed line: $S^{\mathrm{drag}}$.
The circle, diamond, and triangle respectively denote the theoretical
value of $S^{\mathrm{tot}}$, $S^{\mathrm{diff}}$, and $S^{\mathrm{drag}}$
taken from Ref.~\citep{Zhou:2015}. Squares: Experimental data~\citep{Geballe:1955}.\label{fig:2}}
\end{figure}

We start by examining the effect of spin-orbit coupling (SOC) on the Seebeck coefficient for holes,
as no such report is available in the literature.
 Our results for $\mathrm{\mathit{S}^{tot}}$ in hole-doped bulk silicon with and without SOC,
together with the corresponding contributions from the diffusion and drag parts, are shown in Fig. \ref{fig:2} (panels~a~and~b) as a function of carrier
concentration at 300~K. Taking SOC into account leads to a decrease of $\mathrm{\mathit{S}^{tot}}$ at all concentrations, significantly improving the agreement with the available experimental data~\citep{Geballe:1955}.
Turning now to the analysis of $\mathrm{\mathit{S}^{diff}}$ and $\mathrm{\mathit{S}^{drag}}$ contributions,
 the absolute value of $\mathrm{\mathit{S}^{diff}}$
(dashed line), as expected, decreases linearly with the increase
of carrier doping~\citep{Zhou:2015,Fiorentini:2016,Protik:2020,Xu:2021,Protik:2022}, and is not
affected by the presence/absence of SOC. In wide contrast,
$\mathrm{\mathit{S}^{drag}}$ (dot-dashed line) remains nearly independent
of the carrier concentration and is found to be strongly affected by SOC.
Indeed, we find that the $\mathrm{\mathit{S}^{drag}}$ contribution is reduced
by $30\%$ when SOC interaction is accounted for. 
As the effect of SOC on the electron-phonon
matrix elements in Si is found to be weak for phonons contributing to phonon drag (see SM),
this reduction is explained
by the change  of the band structure around the top of the valence band (VB) induced by SOC~\citep{Ponce:2018}, which,
in turn, affects the number of allowed electron-phonon interactions contributing to the phonon drag
(the Dirac distribution in Eq.~\ref{eq:2}). Indeed, Ponc\'e \textit{et al.}~\citep{Ponce:2018} pointed out the
improvement of the calculated hole effective masses in the valence band of silicon with SOC.
We show that this is crucial also for the drag Seebeck coefficient for holes.

We now turn to the role of dimensionality, size, and
direction in governing the drag Seebeck coefficient of silicon nanostructures. Our theoretical results (Fig. \ref{fig:1})
show the effect of size reduction on $\mathrm{\mathit{S}^{drag}}$ for monocrystalline
intrinsic samples of different geometries at 300~K.
These results have been obtained for $p=0$ in Eq. 4 (completely diffusive boundary) and thus, should
be regarded as the lowest threshold value of $\mathrm{\mathit{S}^{drag}}$.
Our calculations show that $\mathrm{\mathit{S}^{drag}}$ is almost
size-independent down to $L^{\mathrm{Cas}}\sim100$~$\mu\mathrm{m}$,
and then decreases monotonically with the decrease in $L^{\mathrm{Cas}}$ for all
dimensionalities and heat transport directions.

However, one can observe a different rate of decrease of $\mathrm{\mathit{S}^{drag}}$
for in- and out-of-plane directions. For both hole- and electron-phonon drag effects, $\mathrm{\mathit{S}^{drag}}$
along the out-of-plane direction of thin films and nanowires behaves
as in the case of the isotropic boundary, and is quenched almost completely
as $L^{\mathrm{Cas}}$ approaches $\sim100$ nm. At the same time,
$\mathrm{\mathit{S}^{drag}}$ decreases at a slower rate
along the in-plane direction of thin films and nanowires than in the out-of-plane one, reflecting the fact
that the phonons are scattered less frequently by boundaries when
traveling in the in-plane direction. Our study shows that for both low hole and electron doping, a silicon thin film (nanowire) of thickness (diameter)
$100$ nm can still preserve more than $20\%$ ($10\%$) of the bulk
$\mathrm{\mathit{S}^{drag}}$ when measured along the in-plane direction.

It must be noted that 
boundary scattering  strongly affects the
phonons with long mean free paths which contribute mostly to phonon drag.
In that respect, the effect of boundary scattering on the phonon drag differs from that of alloying, which
was recently studied in Ref.~\citep{Xu:2021} for Si-Ge alloys.
Indeed, it was shown in Ref.~\citep{Xu:2021} that 
 mass disorder in alloys scatters mostly high-frequency phonons, 
affecting the phonons with  long mean free paths in lesser extent, in contrast to boundary scattering.

\begin{figure}[t!]
\includegraphics[width=8.7cm]{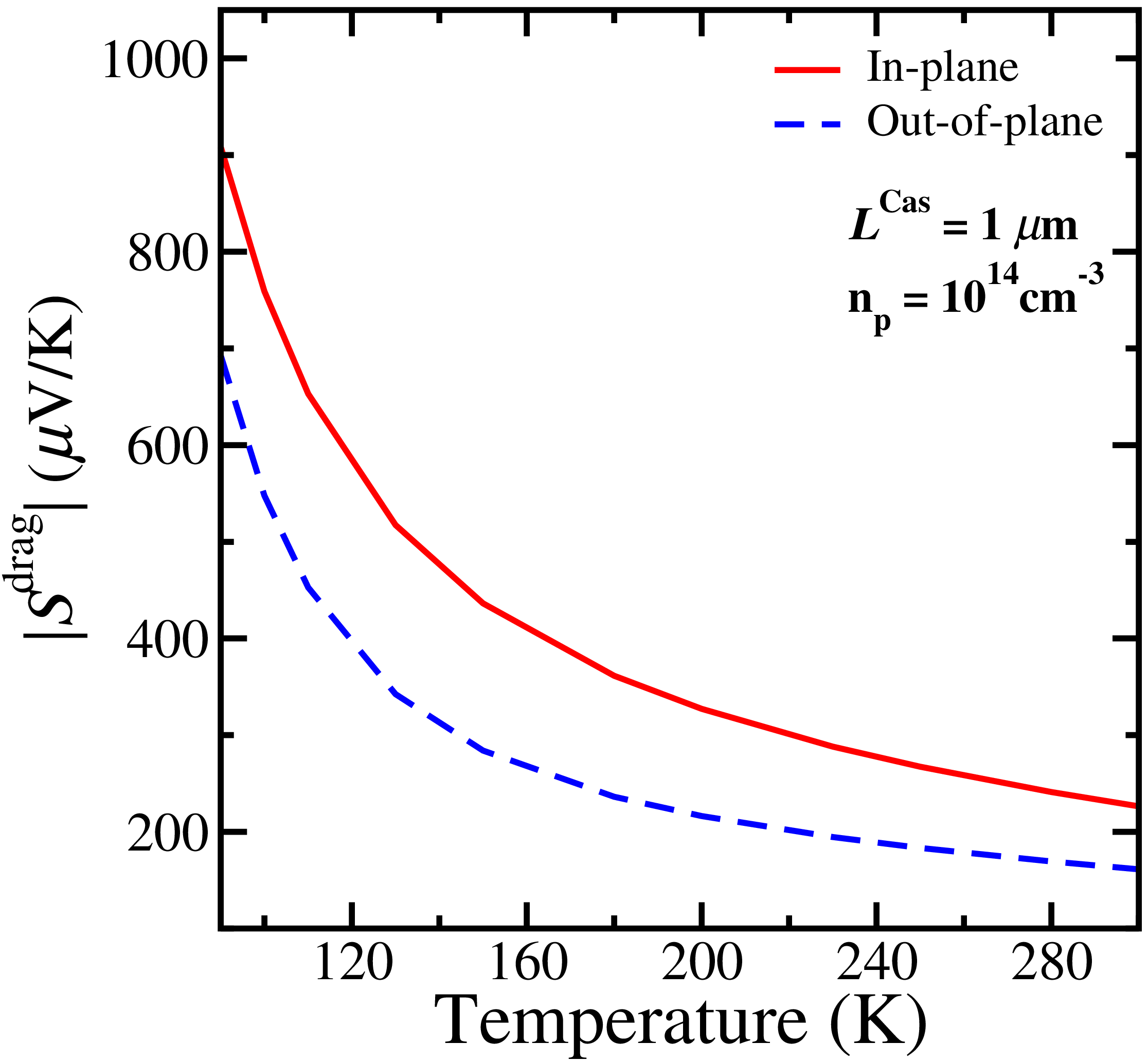}
\caption{Variation of hole-phonon drag Seebeck coefficient of silicon thin film
(with $L^{\mathrm{Cas}}=1$~$\mu$m and doping $10^{14}~\mathrm{cm^{-3}}$) as a function of temperature.
The solid and dashed lines represent $\mathrm{\mathit{S}^{drag}}$ along the in-plane and
out-of-plane directions of the thin film, respectively.\label{fig:3}}
\end{figure}

\begin{figure}[t]
\includegraphics[width=8.7cm]{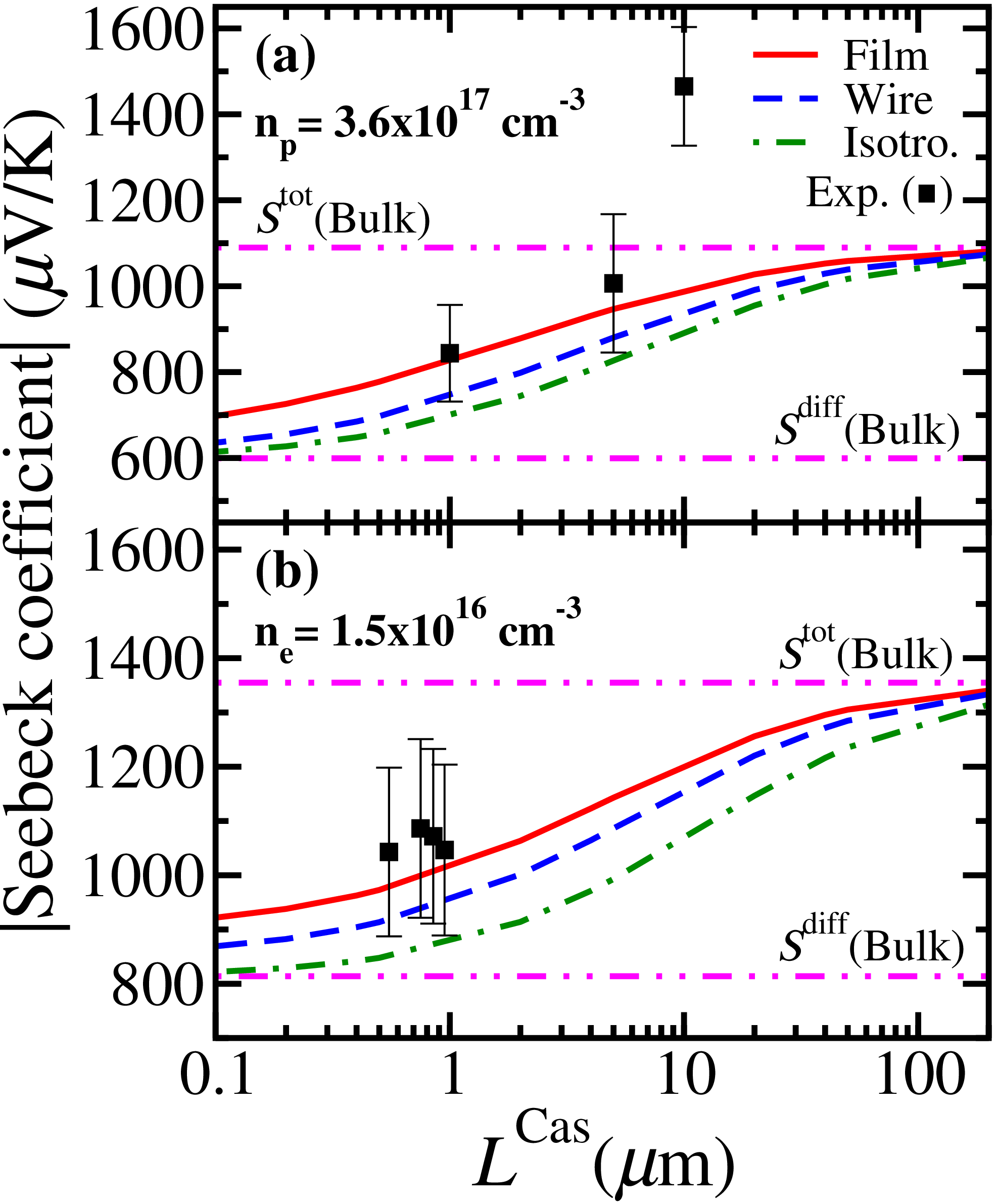}
\caption{Variation of the total Seebeck coefficient ($\mathrm{\mathit{S}^{tot}}$)
of (a) hole-doped ($3.6\times10^{17}~\mathrm{cm^{-3}}$) and (b) electron-doped ($1.5\times10^{16}~\mathrm{cm^{-3}}$) silicon nanowires as a function of the diameter ($L^{\mathrm{Cas}}$) at 300~K. Squares: Experimental data from Ref.~\citep{Fauziah:2020} for holes and Ref.~\citep{Bennett:2015} for electrons.\label{fig:4}}
\end{figure}

Turning to the comparison with previous experimental works, our results do not entirely confirm the conclusions of
Ref.~\citep{Sadhu:2015}, in which the phonon
drag Seebeck coefficient is reported to be quenched completely in silicon nanowires
of diameter smaller that $100$ nm. Rather, our results are found to be compatible
with Refs.~\citep{Fauziah:2020,Bennett:2015}, in which it has been found
that a part of $\mathrm{\mathit{S}^{drag}}$ survives in nanostructures.
One must note here that
the remaining drag contribution which we predict
for silicon nanostructures with $L^{\mathrm{Cas}}=100$ nm would be
within the error bar of the experimental measurements of Sadhu \textit{et
al.}~\citep{Sadhu:2015} at moderate doping ($3\times10^{18}~\mathrm{cm^{-3}}$).

We next predict the temperature dependence of $\mathrm{\mathit{S}^{drag}}$ in silicon nanostructures (Fig. \ref{fig:3}). For hole-doped silicon thin film ($10^{14}~\mathrm{cm^{-3}}$) with $L^{\mathrm{Cas}}=1$~$\mu$m, the $\mathrm{\mathit{S}^{drag}}$
contribution is found to grow with decreasing temperature.
The in-plane temperature dependence is found similar to the out-of-plane one.
This behaviour is consistent with the increase of $\mathrm{\mathit{S}^{tot}}$ of bulk
silicon~\citep{Geballe:1955,Zhou:2015} and is predicted here for the first time for  nanostructures.

Finally, in Fig. \ref{fig:4}, we compare our
theoretical results with recent experiments available in literature:
Ref.~\citep{Fauziah:2020} for hole doped (panel~a)
and Ref.~\citep{Bennett:2015} for electron doped (panel~b) silicon nanowires.
These experiments were performed at room temperature and moderate doping concentrations ($3.6\times10^{17}~\mathrm{cm^{-3}}$ for hole doped and
$1.5\times10^{16}~\mathrm{cm^{-3}}$ for electron doped samples).
As mentioned in the introduction,  nanostructuring can induce several competing effects on the experimentally measured Seebeck coefficient~\citep{Hu:2020}.  
Apart from the phonon drag reduction,  other effects discussed in literature are:  energy filtering  (creation of energy barriers due to extended defects), 
changes in the band structure due to confinement, 
changes in band structure due to fabrication and/or measurement procedures~\citep{Hu:2020} as
well as additional 1D-like phonon transport in case of nanowires~\citep{Boukai:2008,Zhou:2015}.
  The latter phenomena may lead to the increase of the Seebeck coefficient. 
Although the study of those effects
is beyond the scope of our formalism, the comparison between our calculated results and experiments allows to gain insight into the 
relative role of phonon drag with respect to other effects. Indeed, one can see in panel~a of Fig.~\ref{fig:4} that in agreement
with our calculations, the experimentally measured hole Seebeck coefficient in Ref.~\citep{Fauziah:2020} was found to decrease with the decreasing nanostructure size, which is a clear indication that phonon drag contribution is still present for the nanostructure sizes under study.
At the same time, some other effect is
clearly playing a role for the Seebeck coefficient measured in Ref.~\citep{Fauziah:2020}, which is demonstrated by the
fact that the experimental Seebeck value at 10~$\mu$m exceeds the bulk value at the same doping concentration.

For the case of electrons~\citep{Bennett:2015}, the Seebeck coefficients which were measured
for defect-free nanowires with diameters 0.6-1~$\mu$m are found in good agreement with our calculated data,
demonstrating the effectiveness of our theoretical scheme (Fig. ~\ref{fig:4}, panel~b).
We note that the isotropic boundary model underestimates $\mathrm{\mathit{S}^{tot}}$
for both cases of electrons and holes.

In conclusion, in this work we have provided a detailed \textit{ab initio} study of the effect of the dimensionality, size, and heat-transport direction on the phonon drag Seebeck coefficient in silicon nanostructures, accounting both for the anisotropy of the boundary scattering and for the spin-orbit coupling.
Inclusion of the latter is shown to be mandatory to obtain a predictive description of the hole-phonon scattering.
The implementation of the phonon drag term in combination with anisotropic
scattering of phonons by nanostructure boundaries
turns out to be crucial to evaluate the transport-direction-dependent  out-of-equilibrium  phonon populations in silicon nanostructures, and predict the phonon-drag contribution to the Seebeck coefficient. In particular, we have shown that even if
phonon drag contribution is strongly reduced by nanostructuring, a silicon thin film (nanowire) of thickness (diameter)
$100$ nm can still preserve, at 300 K, more than $20\%$ ($10\%$) of the bulk
$\mathrm{\mathit{S}^{drag}}$ when measured along the in-plane direction, for both electrons and holes.
Our findings for silicon nanowires support the conclusion of the recent experimental work of Ref.~\citep{Fauziah:2020} concerning the impact of phonon-boundary scattering on the hole Seebeck coefficient of silicon nanowires and show an excellent agreement with the electron Seebeck coefficient measured in Ref.~\citep{Bennett:2015}.
At the same time, the remaining drag contribution which we predict
for silicon nanostructures with $L^{\mathrm{Cas}}=100$ nm would be
within the error bar of the experimental measurements of Ref.~\citep{Sadhu:2015} at moderate doping. Therefore, our results allow to resolve the apparent contradiction in previous literature.
Furthermore, we also predict that even if the remaining  contribution of $\mathrm{\mathit{S}^{drag}}$  at 300 K  is
relatively small in silicon nanostructures, a remarkable increase of the phonon drag contribution is to be expected at low temperatures.

\begin{acknowledgments}

Calculations have been performed with the Quantum ESPRESSO computational package~\cite{Giannozzi:2017}, the EPW code~\cite{Ponce:2016}, the D3Q code~\cite{Paulatto:2013,Fugallo:2013} and
 the Wannier90 code~\cite{Pizzi:2020}.
This work has been granted access to HPC resources by he French HPC centers GENCI-IDRIS, GENCI-CINES and GENCI-TGCC (Project 2210) and by the Ecole Polytechnique through the 3L-HPC project.
 Financial supports from the ANR (PLACHO project ANR-21-CE50-0008, Macacqu flagship Labex Nanosaclay ANR-10-LABX-0035),
from the DIM SIRTEQ, from the CNRS-CEA program "Basic reseach for energy" are gratefully acknowledged.

We acknowlege useful discussions with Dr. Natalio Mingo and Dr. Samuel Ponc\'e, as well as the contribution
of Dr. Gaston Kane on the preliminary stage of the project.

\end{acknowledgments}

\bibliographystyle{apsrev4-1}

\bibliography{../../SVN_Papier/biblio/mybiblio}

\end{document}